# Colloidal CuFeS$_2$ Nanocrystals: Intermediate Fe d-Band Leads to High Photothermal Conversion Efficiency

**Sandeep Ghosh**[†,‡,*], **Tommaso Avellini**[⊥,‡], **Alessia Petrelli**[⊥,‡], **Ilka Kriegel**[†,‡], **Roberto Gaspari**[∥,#,‡], **Guilherme Almeida**[†,Δ], **Giovanni Bertoni**[†,§], **Andrea Cavalli**[∥,#], **Francesco, Scotognella,**[Ω] **Teresa Pellegrino**[⊥,*] **and Liberato Manna**[†,*]

[†]Department of Nanochemistry, Istituto Italiano di Tecnologia (IIT), via Morego 30, 16163 Genova, Italy
[#]CONCEPT Lab, Istituto Italiano di Tecnologia (IIT), via Morego 30, 16163 Genova, Italy
[§]IMEM-CNR, Parco Area delle Scienze 37/A, I-43124 Parma, Italy
[Δ]Dipartimento di chimica e chimica industriale, Università degli Studi di Genova, Via Dodecaneso, 31, 16146, Genova, Italy
[∥]CompuNet, Istituto Italiano di Tecnologia (IIT), via Morego 30, 16163 Genova, Italy
[⊥]Department of Drug Discovery and Development, Istituto Italiano di Tecnologia (IIT), via Morego 30, 16163 Genova, Italy
[Ω] Dipartimento di Fisica, Politecnico di Milano, Piazza Leonardo da Vinci 32, 20133 Milano, Italy

**ABSTRACT:** We describe the colloidal hot-injection synthesis of phase-pure nanocrystals (NCs) of a highly abundant mineral, chalcopyrite (CuFeS$_2$). Absorption bands centered at around 480 and 950 nm, spanning almost the entire visible and near infrared regions, encompass their optical extinction characteristics. These peaks are ascribable to electronic transitions from the valence band (VB) to the empty intermediate band (IB), located in the fundamental gap and mainly composed of Fe 3d orbitals. Laser-irradiation (at 808 nm) of an aqueous suspension of CuFeS$_2$ NCs exhibited significant heating, with a photothermal conversion efficiency of 49%. Such efficient heating is ascribable to the carrier relaxation within the broad IB band (owing to the indirect VB-IB gap), as corroborated by transient absorption measurements. The intense absorption and high photothermal transduction efficiency (PTE) of these NCs in the so-called biological window (650-900 nm) makes them suitable for photothermal therapy as demonstrated by tumor cell annihilation upon laser irradiation. The otherwise harmless nature of these NCs in dark conditions was confirmed by *in vitro* toxicity tests on two different cell lines. The presence of the deep Fe levels constituting the IB is the origin of such enhanced PTE, which can be used to design other high performing NC photothermal agents.

INTRODUCTION

Chalcopyrite (CuFeS$_2$), a mineral with golden luster, is an n-type semiconductor that exhibits large thermoelectric power,[1] rectification,[2] antiferromagnetism with Néel temperature of 823 K,[1] and an unusually low optical band gap of 0.53 eV.[3] Chalcopyrite crystallizes in a substitutional analogous structure to that of carbon, which means that all the atoms are tetrahedrally co-ordinated in the lattice, belonging to the tetragonal space group I$\bar{4}$2d.[4] Despite being composed of earth abundant elements like Cu and Fe, there are very few comprehensive studies on NCs of this Cu-based semiconductor in the literature.[5-6] Silvester *et al.* have reported the hydrothermal synthesis of optically transparent colloidal sols of chalcopyrite NCs through the controlled conversion of amorphous iron (III) oxide particles,[7] with optical characteristics fairly matching to those of bulk.[8] Wang *et al.*[9] and Liang *et al.*[10] have described colloidal approaches to synthesize CuFeS$_2$ NCs, although the optical properties and electronic band structure have not been explored in much detail. For instance, Liang *et al.* reported a band gap of 1.2 eV for the CuFeS$_2$ NCs which is reminiscent of the plasmonic absorption of Cu$_{2-x}$S NCs,[11] suggesting that their samples probably contained a fraction of Cu$_{2-x}$S NCs.

Here, we report a synthesis protocol that delivers nearly monodisperse CuFeS$_2$ NCs that are free from contaminant phases, since to date there have been no consistent reports on phase pure, stable and monodisperse NCs for this material. The optical reflectivity spectrum of a single crystal of CuFeS$_2$ is characterized by strong absorption bands at around 1.0 and 2.1 eV and an absorption edge at around 3.2 eV,[8] as also present in the optical absorption spectra from solutions of the NCs prepared by us. These absorption characteristics arise from an intermediate band (IB) in the fundamental gap, formed predominantly by empty Fe 3d orbitals. Electronic transitions to the IB from the valence band (VB) would explain the lower-energy absorption bands, while the edge at 3.2 eV can be assigned to the VB to conduction band (CB) transitions.[12-14] The top of the VB and the bottom of the IB are found to be separated by an indirect gap of 0.53 eV.[3]

Iron is known to introduce deep intermediate trap states in the fundamental gap of most photovoltaic semiconductors and hence is meticulously avoided in commercial silicon photovoltaics, since it drastically affects the solar cell efficiency by introducing numerous recombination centers.[15] In CuFeS$_2$, iron can be considered to play a very similar role since the broad IB is mainly composed of Fe 3d orbitals, and is generally absent in other chalcopyrite type semiconductors like CuGaS$_2$ and CuAlS$_2$.[16] This band structure is indeed supported by density functional theory (DFT) calculations per-performed by us and reported here, which yields an indirect VB-IB gap of 0.63 eV, in reasonable agreement with the bulk experimental value. The presence of an indirect VB-IB gap means that the electrons photo-excited to the IB will relax through thermal non-emissive pathways. Indeed, transient absorption (TA) measurements on a solution of the NCs synthesized by us are consistent with thermalization of the photoexcited carriers within the broad intermediate band (IB), followed by the interband relaxation from the IB to the VB (slower).

Here, we take advantage of this unique band structure to demonstrate that $CuFeS_2$ NCs can be employed as efficient light-to-heat converters in the biological window (650-900 nm, *vide infra*), owing to its low optical gap. Likewise, the as-synthesized NCs, coated with a commercially available amphiphilic polymer that made them dispersible in aqueous solutions, exhibited significant heating in water upon exposure to a laser at 808 nm, with photothermal transduction efficiency (PTE) of 49%. This notion is further corroborated by the high molar attenuation coefficient ($\varepsilon = 5.2 \times 10^6$ $M^{-1}cm^{-1}$) of the NCs at this irradiation wavelength. We further demonstrate that these characteristics make the NCs of chalcopyrite ($CuFeS_2$) efficient agents for photothermal tumor therapy. Laser-induced ablation of cancer cells or photothermal therapy (PT) has been gaining increasing importance in recent years as a viable method of cancer treatment,[17] mostly due to its much less invasive nature.[18] In order to be fit for PT applications, the sensitizers must: i) exhibit strong absorption in the near-infrared (NIR) region, especially in the 650-900 nm range (the so-called biological window, where light extinction by hemoglobin and water is minimal);[19] (ii) efficiently convert the absorbed NIR radiation into heat. In metallic or heavily doped systems with a high density of free carriers, like plasmonic nanomaterials of Au [20-23] and copper chalcogenides,[24-28] heat release occurs due to the excitation of localized surface plasmon resonances. Immediately after excitation, the hot carriers thermalize and cool through the emission of phonons by interacting with the crystal lattice of the nanoparticle. This leads to heating of the nanoparticle, which is then exploited for PT. Heat generation is limited to the region of the plasmon resonance, which itself depends on the carrier density, the size and the shape of the nanoparticle.[29] Recently, various semiconductor NCs, among which Ge,[30] Si,[31] $WS_2$,[32-33] $W_{18}O_{49}$,[34] $Bi_2S_3$,[35] $Nd:LaF_3$,[36] and $NdVO_4$ [37] have also been explored as PT agents, although in most of these reports, the mechanism and the essential requirements for light-to-heat conversion have not been thoroughly discussed. For semiconductors in general, carriers with excess energy (above the band edge) relax by emission of phonons, which ultimately leads to heating of the material.[38] For photothermal applications in the NIR this is however limited to semiconductors with a sufficiently low band gap.

The $CuFeS_2$ NCs synthesized by us fulfill both the criteria of efficient NIR light to heat conversion and strong absorption in the NIR region. However, it was necessary to demonstrate the therapeutic ability of these NCs on live cells upon laser irradiation and check for their cytotoxicity under radiationless conditions. An ideal photothermal agent should not be toxic to live cells in dark conditions. To this effect, viability tests on two different cell lines were undertaken to ascertain the cytotoxicity of $CuFeS_2$ NCs under radiationless conditions. The cells could tolerate a NC dose of up to 50 ppm of copper content for 24 h in the dark. However, upon prolonging the exposure to 72 h in the dark, even lower NC doses (less than 10 ppm in Cu) were found to be cytotoxic. The therapeutic ability of the NCs were then demonstrated through annihilation of the NC-imbibed-cells under laser irradiation (808 nm, in the biological window), even at NC doses far lower than the toxic limit. A NC dose of around 3 ppm (in Cu) was identified as the therapeutic limit i.e. this dose is efficient in annihilating cells under laser irradiation but is otherwise harmless in the dark. The efficient light to heat conversion (PTE: 49%), strong NIR absorption ($\varepsilon = 5.2 \times 10^6$ $M^{-1}cm^{-1}$ at 808 nm) and small size (21 nm hydrodynamic diameter, ensures longer bloodstream circulation time) demonstrate that chalcopyrite NCs can be employed as photothermal agents.

EXPERIMENTAL SECTION

**Materials**. Copper (I) iodide (CuI, ≥99.5%), iron (III) acetylacetonate ($Fe(acac)_3$, ≥99.9%), 1-octadecene (ODE, technical grade, 90%), oleylamine (OLAM, technical grade, 70%) and 1-dodecanethiol (DT, ≥98%) were purchased from Sigma-Aldrich. Tri-n-octylphosphine (TOP, 97%), was procured from Strem Chemicals. α-methoxy-ω-mercapto polyethylene glycol (SH-PEG-$OCH_3$, MW: 2000 g $mol^{-1}$) was purchased from Rapp Polymere. Anhydrous chloroform, methanol and isopropanol were purchased from Carlo Erba reagents. All chemicals were used as received. A Milli-Q Millipore system was used for the purification of water (resistivity ≥ 18 MΩ·cm).

**Synthesis of $CuFeS_2$ NCs**. All synthesis procedures were undertaken by employing standard Schlenk line techniques assisted by a nitrogen-filled glove box. Typically, a mixture of CuI (190 mg, 1 mmol) and TOP (380 mg (amount adjusted as per chemical assay), 1 mmol) in a 1:1 molar ratio was prepared in ODE (1 ml). The mixture was homogenized, aided by mild heating and stirring, until all the CuI dissolved resulting in a transparent solution. Care has to be exercised as to prepare this solution in an inert atmosphere. This solution was then transferred to a mixture of $Fe(acac)_3$ (353 mg, 1 mmol) and ODE (6 ml) kept in a 25 ml round-bottomed flask. The final mixture was then subjected to evacuation at 80 °C for a period of 1h, letting the mixture dry. Thereafter, the flask was filled with nitrogen and the temperature was raised to 280 °C. A solution of 1 ml DT and 2ml OLAM, separately prepared by degassing the mixture at 80 °C and subsequently heating it under nitrogen flow to 160 °C, was injected at this point. The temperature dropped to 260 °C, and was allowed to recover to 270 °C. The initial dark red solution (color ascribed to $Fe(acac)_3$) turned dark brown within the first few minutes, signifying the formation of the $CuFeS_2$ NCs. The reaction mixture was allowed to stir for 15 min at this temperature and finally the heating mantle was removed to cool the reaction mixture. The final dark purple solution was then washed with chloroform/methanol mixture. The NCs were resuspended in chloroform and subjected to a further centrifugation at 1000-2000 rpm for 1-2 minutes in order to remove larger particles formed during the reaction.

**X-Ray Diffraction (XRD) analysis**. Concentrated solutions of the NC samples were drop-cast (followed by evaporation of the solvent) on a zero background silicon substrate to acquire the XRD patterns, which was performed on a Rigaku SmartLab 9 kW diffractometer with the X-ray source operating at 40 kV and 150 mA. The instrument was equipped with a Cu source and a Göbel mirror (to obtain a parallel beam and suppress the Cu Kβ radiation at 1.392 Å) and was used in the 2θ/ω scan geometry for data acquisition. Phase identification was performed through the assistance of PDXL software.

**Steady state UV-Vis-NIR extinction spectroscopy**. Optical extinction spectra of dilute chloroform dispersions of the NC samples were recorded in quartz cuvettes of 1 cm path-length employing a Varian Cary 5000 UV-Vis-NIR absorption spectrophotometer in the wavelength range of 300-2200 nm.

**Elemental analysis**. The composition and solution concentration of the NCs was ascertained using Inductively Coupled Plasma-Optical Emission Spectroscopy (ICP-OES) analysis performed on aiCAP 6000 spectrometer (ThermoScientific). The samples were digested overnight in aqua regia and diluted to a known amount before the measurements.

**Dynamic Light Scattering (DLS) measurements**. The hydrodynamic diameter of the NCs was determined by DLS measurements on a Malvern Zetasizer (Nano Series, Nano ZS) instrument. The scattered intensity was collected at 173° back scattered geometry with a 632 nm laser source. For each sample, 3 measurements were taken with 10 to 20 acquisitions in each case.

**Transmission Electron Microscopy (TEM) analysis**. BFTEM images and SAED patterns were acquired on samples prepared by drop-casting a concentrated solution of NCs on carbon-coated 200 mesh copper grids, using a JEOL JEM-1011 microscope (W filament) operated at 100 kV accelerating voltage. HRTEM, HAADF, and EDS analyses were performed on a JEOL JEM-2200FS microscope equipped with a Schottky emitter at 200 kV, a CEOS image corrector allowing for an information limit of 0.8 Å, and an in-column energy filter ($\Omega$-type). The chemical compositions of the NCs were determined by EDS using a JEOL JED-2300 Si(Li) detector. The NC suspensions were deposited onto ultrathin carbon coated Au grids and the measurements were carried out using a holder with a beryllium cup for background reduction in the spectra.

**Computational Details.** Calculations have been performed considering the orthorhombic unit cell, comprising of 4 Fe, 4 Cu and 8 S atoms, using the plane wave code pwscf.[39] For all calculations, we employed ultrasoft pseudopotentials, a Brillouin zone sampling using a 4x4x4 Monkhorst-Pack mesh[40] and a plane wave cutoff of 30 Ry and 240 Ry for the wavefunction and charge density expansion, respectively. The PBE functional[41] and linear response DFT+U[42] was used. The system geometry was relaxed until the largest force on atoms was smaller than $10^{-4}$ Ry/Bohr and the largest stress tensor component was smaller than $3\cdot 10^{-5}$ Ry/Bohr$^3$. Spin polarization was allowed in order to reproduce the antiferromagnetic ground state of the system (see also Fig. S3). Further details can be found in the Supporting Information.

**Transient absorption (TA) spectroscopy.** Ultrafast pump-probe measurements were performed on a Ti: Sapphire chirp pulse amplified source, with maximum output energy of about 1 mJ, 1 kHz repetition rate, central wavelength of 800 nm and pulse duration of about 150 fs. Pump pulses of 490 nm were generated by optical parametric amplification (OPA) in a β-Barium borate (BBO) crystal. White light was generated in a thin sapphire plate in approximately the range of 430 – 700 nm. The detection system was based on a fast optical multi-channel analyzer (OMA) with dechirping algorithm to obtain chirp-free transient transmission spectra measured as the normalized transmission change, $\Delta T/T$. The excitation energy per pulse was kept in the linear regime with a maximum fluence of ~80 μJ/cm$^2$. All measurements were performed at room temperature on sealed samples prepared under nitrogen atmosphere by dispersing the CuFeS$_2$ NCs in chloroform.

**Ligand exchange and water transfer.** The CuFeS$_2$ NCs were transferred to water by replacing the native ligands with hydrophilic thiol terminated polyethylene glycol (PEG) molecules, SH-PEG-OCH$_3$. In a typical exchange procedure, 3 ml of the NC-solution in chloroform (3.0 μM, 12.6 nm tetrahedral edge) was mixed with a solution of SH-PEG-OCH$_3$ in methanol (248 mg dissolved in 15 ml of methanol). The amount of PEG units was roughly determined based on a fixed number of ligand molecules per unit NC surface area (50 ligand molecules per square nm of NC surface, in this case). The mixture was shaken vigorously for 2 h and the PEG coated NCs were then washed with 20 ml of hexane. The concentrated NCs solution was diluted with about 4 ml of methanol and the washing step described above was repeated twice to ensure an efficient removal of the hydrophobic ligands (*e.g.* dodecanethiol/oleylamine). The solvent from the PEG-coated NC solution was then removed under reduced pressure yielding the dried NCs, which produced a clear homogeneous purple solution when dispersed in water. Some large NC aggregates were formed during these manipulations and were removed by syringe filtration (0.2 μm pore size). Finally, the excess of free PEG polymer was removed with two cycles of water dilution/concentration filtration using a Millipore Amicon centrifuge filter (100 KDa MWCO, 2300 rpm, 10 min).

**Photothermal transduction efficiency (PTE) calculation.** The PTE of the PEG-coated NCs was determined according to the protocol of Roper *et al.*[43] A quartz cuvette, containing an aqueous solution of NCs, was placed inside a vacuum chamber and irradiated with a 808 nm continuous laser (RTLMDL-808-5W, Roithner Laser Technik) while the temperature was monitored using a thermo-probe (FOT Lab Kit equipped with Fluoroptic probe, Luxtron). The laser was switched off when the solution temperature reached a plateau and the temperature *vs.* time profile of the solution was recorded. As a control, the same heating-cooling cycle was performed for a quartz cuvette filled with the same volume of deionized water. From the cooling profiles, the photothermal transduction efficiency was determined (see section S4 of the SI for details).

**Cell culture.** Two cell lines were employed in our experiments. Human epithelial carcinoma cells (HeLa, ATCC, UK) were cultured in Dulbecco's Modified Eagle Medium (DMEM, Gibco, UK) and human ovarian cancer cells (IGROV 1, ATCC, UK) were cultured in Roswell Park Memorial Institute Medium (RPMI, Gibco,UK) both supplemented with 10% Inactivated Fetal Bovine Serum (FBS), 1% Penicillin Streptomycin (PS) and 1% Glutamine at 37 °C, in 95% humidity and 5% CO$_2$. Cells were split every 3-4 days before reaching 80 % confluence.

**Stability of PEG-coated NCs in physiological conditions.** NCs at a dose of 10 ppm Cu were dispersed in complete DMEM medium and kept at 37 °C, at 95% humidity and 5% CO$_2$ for different intervals of time. The hydrodynamic diameter was monitored over time for the first 24 h by DLS.

**Cell viability tests.** The NC doses to add to the cell dish were standardized against the copper content of the NCs. The PrestoBlue (PB) assay was performed according to the manufacturer's protocol (Invitrogen, Carlsbad, CA, USA). Briefly, 10$^4$ HeLa or IGROV 1 cells/well were seeded in a 96 multi-well plate and allowed to adhere to the well dish. After 24 h of incubation, NCs at increasing doses (measured as the Cu concentration, [Cu], ranging from 1 to 50 ppm) were then added to the media and incubated for additional 24 and 72 h at 37 ° C. The PB reagent was then added directly to the cells in the culture medium and incubated for additional 60 minutes at 37 °C. The cell viability was determined at this stage by recording the absorbance for each well at 560 and 600 nm. The absorbance ratios (A560/A600 nm) for each well were normalized with respect to the absorbance ratio of control cells wells which were not treated with $CuFeS_2$ NCs.

***In-vitro* photothermal ablation experiments.** $2 \times 10^5$ HeLa cells/well were seeded in a 6-well plate and allowed to adhere for 24 hours. NCs (at Cu content ranging from 3 to 20 ppm) were then added to the cell media and incubated for an additional 24 h at 37 °C. The cells were then rinsed with phosphate-buffered saline (PBS) to remove the non-internalized NCs and collected by centrifugation, after being detached by trypsin treatment. The cells were then transferred to a 2 ml Eppendorf tube so that $10^6$ cells were present in 0.5 ml of complete DMEM media, so that they were ready for the irradiation experiments. Irradiation experiments were performed using an 808 nm continuous laser set up (RLTMDL-808-5W, Roithner Laser Technik, power adjusted to 1.14 W, spot size 0.36 $cm^2$ and power density to 3.1 W $cm^{-2}$) equipped with a controlled temperature chamber to fix the temperature of the sample holder at 37.0 ± 0.1 °C. The irradiation time for each experiment was set at 13 min and the temperature was monitored during the treatment using a thermo-camera (OPTRIS, PI230). Temperature profiles were recorded on cell suspensions and also on solutions containing only NCs (for comparison), at NC concentrations corresponding to those administrated to the cells. To this aim, NC solutions containing 3, 5, 10, 20 ppm of Cu, respectively, were dispersed in a final volume of 0.5 ml of complete media (DMEM) and irradiated at 808 nm (1.14W, 3.1 $Wcm^{-2}$) for 13 min. To test the toxic effects of the irradiation on the cells, the radiation treated HeLa cells were re-seeded in a 24 well plate at a density of $2 \times 10^5$ /well and their viability was analyzed by PB assay after 24 and 72 h of incubation at 37 °C. The cell viability was also assessed through fluorescence imaging. The LIVE/DEAD Viability Assay was performed according to the manufacturer's protocol (Thermo Fisher Scientific). Briefly, after NIR irradiation experiments, $2 \times 10^5$ HeLa cells/well were seeded in a 6-well plate and allowed to adhere for 5 hours, after which they were washed gently in PBS and incubated in dark conditions for 45 minutes with 100 μl per well of LIVE/DEAD reagents. The LIVE/DEAD reagent is composed of 20 μl of the supplied 2 mM Ethidium Homodimer (EthD-1) stock solution and 5 μl of 4 mM calcein AM dye of the supplied stock solution in 10 mL of sterile PBS. Green fluorescence for calcein AM corresponding to living cells and red fluorescence for EthD-1 corresponding to dead cells was detected using a Nikon A1 confocal microscope.

RESULTS AND DISCUSSION

**Synthesis and characterization of the NCs.** In our synthesis, we prepared phase pure NCs through a colloidal approach wherein copper (I) iodide (CuI) and iron (III) acetylacetonate were employed as the respective Cu and Fe precursors and 1-dodecanethiol (DT) as the S-source which was delivered via hot-injection, in the presence of oleylamine as a capping agent. CuI was first dissolved in TOP in the presence of ODE since it is otherwise insoluble in ODE and also since TOP is known to favor the formation of the ternary chalcopyrite phase.[44] CuI was chosen as the Cu-precursor since other copper halides like CuCl and CuBr led to the formation of Cu-S impurity phases. This happens due to the increased reactivity of $Cu^+$ ions (soft acid) towards the thiol moiety (soft base), when present in association with harder bases like $Cl^-$ or $Br^-$. $I^-$ being a much softer base reduces the rate of $Cu^+$ reaction with DT and balances the reactivity of the two metal precursors, leading to phase purity.[45] The reaction led to dark purple colloidal dispersions of chalcopyrite NCs, as shown in Figure 1a. The phase purity of the as-synthesized $CuFeS_2$ NCs was ascertained through XRD and azimuthally integrated SAED patterns, as shown in Figure 1b, which confirm the tetragonal ($I\bar{4}2d$) crystal structure as reported for the bulk structure by Knight *et al.*[4]

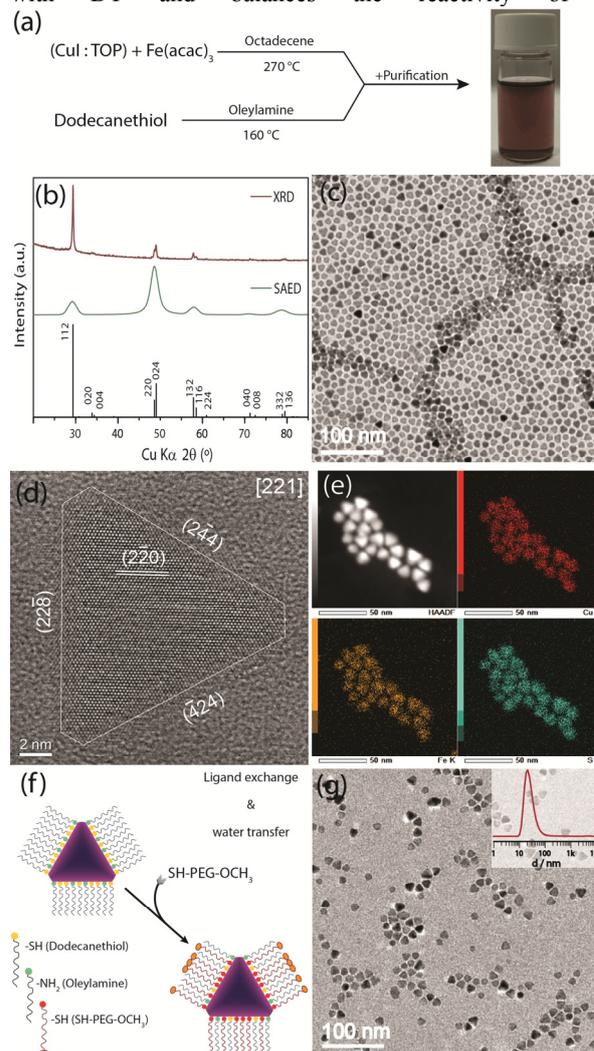

**Figure 1.** (a) Reaction scheme leading to the purple colored solution of $CuFeS_2$ NCs in chloroform (vial photograph). (b) XRD pattern and azimuthally integrated SAED patterns for $CuFeS_2$ NCs, in reference to the database powder XRD pattern ($I\bar{4}2d$, ICDD #96-901-5235). (c) Overview BFTEM image of a set of NCs. (d) HRTEM image of a single NC viewed along the [221] direction. (e) HAADF-STEM image and corresponding STEM-EDS maps showing the distribution of the constituent elements. (f) Ligand exchange scheme for transferring the NCs to aqueous media using an amphiphilic polymer, and (g) BFTEM image of $CuFeS_2$ NCs deposited from an aqueous solution with the hydrodynamic diameter (by volume), shown as an inset.

The chalcopyrite phase was discernible within the first few minutes after the injection of the sulphur source (DT-OLAM mixture) into the Cu-Fe pre-injection salt solution (degassed

CuI-TOP-Fe(acac)$_3$-ODE mixture), as confirmed by XRD patterns of the aliquots taken at different time intervals (Figure S1). Unlike Silvester *et al.*,[7] where a controlled conversion of amorphous iron (III) oxide particles into CuFeS$_2$ was demonstrated, we have achieved a direct synthesis protocol. Monitoring the growth of the NCs revealed some interesting characteristics: the NCs exhibited a quasi-cuboctahedral shape within 1 min after the injection (average size 10.5 ± 0.6 nm; BFTEM images in Figure S1), which slowly ripened into the eventual pyramidal shape in the course of 15 min after injection, as shown in the representative overview BFTEM image in Figure 1c. The latter stages (*e.g.* 25, 30 and 45 mins; see Figure S1) were characterized by further ripening accompanied with truncation of the pyramid vertices leading to increased size dispersity. The unambiguous phase purity, however, could be ascertained only after a growth time of 15 min.

The elemental composition of the NCs was determined through ICP-OES analysis. The Cu: Fe molar ratio of the NCs formed in the first few minutes (1-5 min) was slightly greater than 1, *i.e.* the NCs were Cu-rich. The desired value for Cu:Fe:S molar ratios of 1:1:2 was achieved after 15 min of growth, which lends further proof to phase purity. The NCs discussed henceforth are those that were allowed to grow for 15 min in the reaction mixture, and were the most monodisperse ones, with an average edge length of 12.9 (±1.1) nm (Figure 1c). An HRTEM image of a single NC viewed along the [221] crystal direction is shown in Figure 1d and the EDS elemental maps in Figure 1e establish the distribution of the constituent elements in the set of NCs shown in the HAADF-STEM image in the upper left panel.

The hydrophobic CuFeS$_2$ NCs, capped with a mixture of oleylamine/dodecanethiol, were transferred to the aqueous media by means of a ligand exchange procedure. The native ligands on the NC surface were exchanged with SH-PEG-OCH$_3$ molecules bearing a thiol (-SH) moiety as anchoring group for the NC surface and a polyethylene glycol chain (PEG) with the methoxy (-OCH$_3$) head group as the hydrophilic unit. The hydrophilic groups (attached to the NC surface and directed outwards) enabled the dispersion of CuFeS$_2$ NCs in aqueous media, as schematically shown in Figure 1f. The PEG-stabilized CuFeS$_2$ NCs (CuFeS$_2$-PEG) retained their morphology and shape upon transferring to water, as is obvious from the BFTEM image shown in Figure 1g. These NCs exhibited a narrow distribution of hydrodynamic diameters (DLS signal maximum at 21nm, inset in Figure 1g) with no other signals at bigger sizes, indicating the absence of aggregates. Thermogravimetric analysis (TGA) estimates an average of 95 PEG ligands per NC with the inorganic CuFeS$_2$ core constituting ≈ 74% of the total mass of an individual NC (see section S2 for details).

**Optical modeling/characterization.** The electronic structure of CuFeS$_2$ was investigated using a DFT approach, through a procedure similar to that presented in earlier works.[12,46] The calculated density of states (DOS) for the corresponding band structure of CuFeS$_2$ (along the Z-Γ-X line; Figure S4b) is shown in Figure 2a. An indirect band gap of 0.63 eV separates the valence band from the lowest energy unoccupied states. These empty states form a narrow band, which is separated from a higher energy conduction band by a secondary energy gap of 1.20 eV. The overall band structure can hence be modeled in terms of a valence (VB), an intermediate (IB) and a conduction band (CB), similar to a previous theoretical work by Hamajima *et al.*[13] The highest VB states are composed mainly of Cu d orbitals and S p orbitals, while the IB band consists by a large amount of states with minority spin, Fe d-orbitals character. The bottom of the CB, on the other hand, has a clear s-character, with almost equal S,

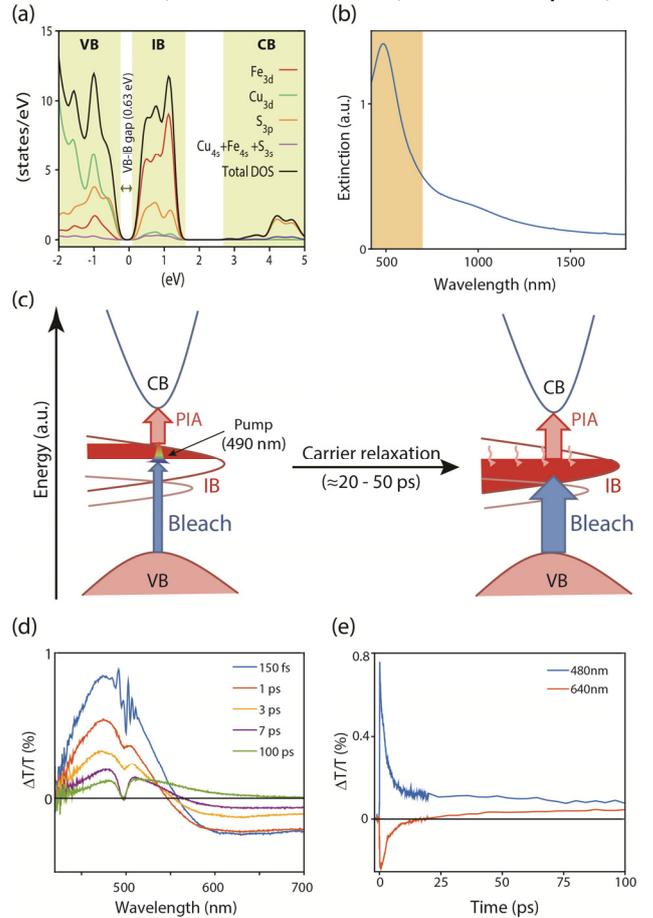

**Figure 2.** (a) Calculated density of states (DOS) for the corresponding electronic band structure. (b) Steady state absorption spectrum of CuFeS$_2$ NCs (shaded area depicting the region of transient absorption measurements). (c) Tentative sketches showing the optical excitations and relaxation processes in CuFeS$_2$. (d) Transient absorption measurements in the range of 0 to 100 ps. The spectrum is contributed by a strong bleach signature (430-550 nm) and a photoinduced absorption (PIA) signal (550–750 nm) at longer wavelengths. A strong broadening of the bleach signal over time is observed. (e) Dynamics at 480 and 640 nm, depicted in blue and orange, respectively.

Fe and Cu contributions. The steady-state UV-Vis-NIR extinction spectrum of a dilute solution of CuFeS$_2$ NCs (Figure 2b) supports this description of the band structure as it features a broad continuous extinction starting from the NIR region (~1800 nm) with an intense maximum at around 480 nm accompanied with a broad shoulder centered at around 950 nm. These two distinct absorption bands, as well as the non-zero absorption in the whole NIR region, are attributed to the electronic transitions from the VB to the IB.

We carried out transient absorption (TA) measurements to elucidate the processes of carrier excitation and relaxation in these NCs. We probed the 430-700 nm spectral range (Figure 2b, shaded region) by pumping at 490 nm into the strong absorption band of CuFeS$_2$, which corresponds to transitions from the VB to the higher intermediate band (IB) (see sketches in Figure 2c). We decided to pump just off the maximum of the transition at 480 nm to minimize the perturbation of the signal by pump scattering. Due to measurement constraints, we focused only on the higher energy VB-IB transition at 480 nm.

However, we are confident that our findings, which are detailed below, should equally apply when investigating the transition from the VB to the lower IB (at 950 nm) possible by an IR pulse. The main findings from these experiments are that the carriers undergo fast relaxation within the IB observed on a time scale of approximately 20-50 ps, before interband relaxation occurs. This fast initial relaxation is ascribed to carrier cooling within the broad IB via the interaction with the lattice and the emission of phonons. This results in the efficient heating of the NC as observed and exploited by us for PT (see further below). Details of the TA measurements are discussed in the following paragraphs.

Pumping at 490 nm into the strong optical absorption band of $CuFeS_2$ results in a bleach around 430-550 nm coinciding with the region of the highest absorption in the steady state spectra, and a broad photo induced absorption signal (PIA) further red. The corresponding TA spectra in the range of 0 to 100 ps are given in Figure 2d. The decay dynamics at the bleach maximum (480 nm, blue curve in Figure 2e) is governed by a two-step decay with a fast component of around 20-50 ps and a slower component of several hundreds of picoseconds. The PIA signal at 640 nm (red curve in Figure 2e) shows a similar decay dynamics and an initial fast rise of less than 1 ps. After about 20-50 ps the negative PIA signal turns to a positive signal. Normalization of the spectra (see Figure S5 in the SI) shows that the spectral shape of the bleach and the PIA is retained during the first 6 ps. Thereafter the spectrum broadens and shifts to the red. After about 20-50 ps the transient spectrum is comprised solely of a bleach signal covering nearly the entire region of investigation, showing a shoulder to the red of the bleach signal.

The excitation and relaxation processes occurring within the first 20-50 ps are depicted schematically in the sketches of Figure 2c, on the band structure of $CuFeS_2$ composed of the VB, a broad, initially empty IB, and the CB. The results have been interpreted in the following way: the pump pulse at 490 nm (rainbow colored pulse in the left sketch) leads to the occupation of states in the IB (red shaded area in Figure 2a), which are narrow in comparison to the width of the IB. An immediate bleach is induced (blue arrow), as these states are occupied and not available for further VB to IB transitions. Simultaneously, a PIA signal is evolving nearly immediately between the IB and the CB (red arrow) due to the temporary occupation of states in the IB. The rather broad PIA signal reflects the manifold of carriers and states involved in this transition within the IB and the CB. The short rise time is a result of the time it takes to fill states in the IB and is within the time resolution of our system. Within the first 20-50 ps the carriers thermalize to the edge of the IB (indicated as small wavy arrows in right sketch of Figure 2a), resulting in a red shift and broadening of the bleach signal (blue arrow). The shoulder to the red of the bleach uncovers the still present negative PIA signal further red, due to IB-CB transitions (red arrow). After cooling of the excited carriers within the first approximately 20-50 ps, a slower interband relaxation is observed (slower component of several hundreds of picoseconds, Figure 2e). Our results confirm that within the first 20-50 ps carrier cooling occurs by transferring the excess energy of the excited carriers in the IB to the lattice of the NC resulting in the heating of the NC, before - on a slower timescale of several hundreds of picoseconds - the interband relaxation takes place. Our results highlight the importance of the unique band structure of $CuFeS_2$ constituting a broad IB for efficient light-to-heat conversion.

**Photothermal conversion and stability studies.** The extinction spectrum of an aqueous solution of PEG coated NCs ($CuFeS_2$-PEG NCs; Figure S6) had features that closely matched those of the as-synthesized NCs in chloroform (Figure 2b). The continuous absorption in the NIR region, particularly within the first biological window (650-900 nm), accompanied with a high molar attenuation coefficient ($\varepsilon = 5.2 \times 10^6$ $M^{-1}cm^{-1}$ at 808 nm; section S5 in SI) makes them potentially suitable candidates for photothermal applications.

The molar attenuation coefficient for the $CuFeS_2$-PEG NCs was comparable to that of other chalcogenide NCs, like CuS nanoplatelets,[47] $Cu_{2-x}$Se NCs[24] and $Cu_9S_5$ NCs[48] and it is at least one order of magnitude higher than that of standard photo-sensitizer dyes used as potential photothermal agents.[49] The molar attenuation coefficient can also be expressed as the per particle absorption cross-section using $C_{abs} = 2303\varepsilon/N_A$ ($N_A$ is the Avogadro constant), considering the fact that scattering contribution to extinction for small NCs can generally be neglected due to their small size with respect to the probing wavelength (808 nm in this case).[50] The per particle absorption cross-section for $CuFeS_2$-PEG NCs at 808 nm is $2 \times 10^{-14}$ $cm^2$.

A remarkable temperature increase was recorded upon irradiation of an aqueous solution of $CuFeS_2$-PEG NCs with an 808 nm laser. The extent of light to heat conversion by a sensitizer is typically expressed as the so-called photothermal transduction efficiency (PTE) η, defined as the fraction of absorbed light that has been converted to heat, and is expressed as follows (as per Roper et al.[43])

$$\eta = \frac{hS(T_{max} - T_{amb}) - Q_{Dis}}{I(1 - 10^{-A_{808}})}$$

where h is the heat transfer coefficient, S is the surface area of the container in which the NC solution was placed, and $T_{max}$ and $T_{amb}$ represent the maximum equilibrium temperature reached by the solution and the ambient temperature, respectively. $Q_{Dis}$ represents the heat dissipated by the solvent and the quartz cuvette itself upon light absorption, $I$ is the incident laser power and $A808$ is the extinction of the sample at 808 nm. As evident from the equation, the PTE represents the amount of heat generated per incident light power and was estimated to be 49 % in our case (see Figure S8b and section S6 of the SI for details on calculations). This value is in line with previous reports on chalcogenide NCs. Overall, we assign this high efficiency to the large molar attenuation coefficient of these NCs and to the thermalization of carriers that are excited from the VB to the IB. The presence of numerous direct VB-IB transitions at this energy difference (808 nm, 1.53 eV), as evident from the electronic band structure shown in Figure S4b, contributes to the high molar attenuation coefficient. Additionally, the absence of photoluminescence in our NCs, both in those as-synthesized and after water transfer, and which can be due to various reasons (presence of trap or defect states, indirect gap between the top of the VB and the bottom of the IB), indicates that even this inter-gap relaxation must occur primarily by thermal processes.

It was important to evaluate the stability of $CuFeS_2$-PEG NCs in physiological conditions, which in turn can affect the photo-thermal performance of the NCs under the laser. Accordingly, a stability test was performed wherein NC solutions with total Cu concentration (henceforth indicated as [Cu]) of 10 ppm were incubated in a complete cell culture medium (DMEM) at 37 °C for 24 h and the time evolution of hydrodynamic diameter was monitored through DLS.

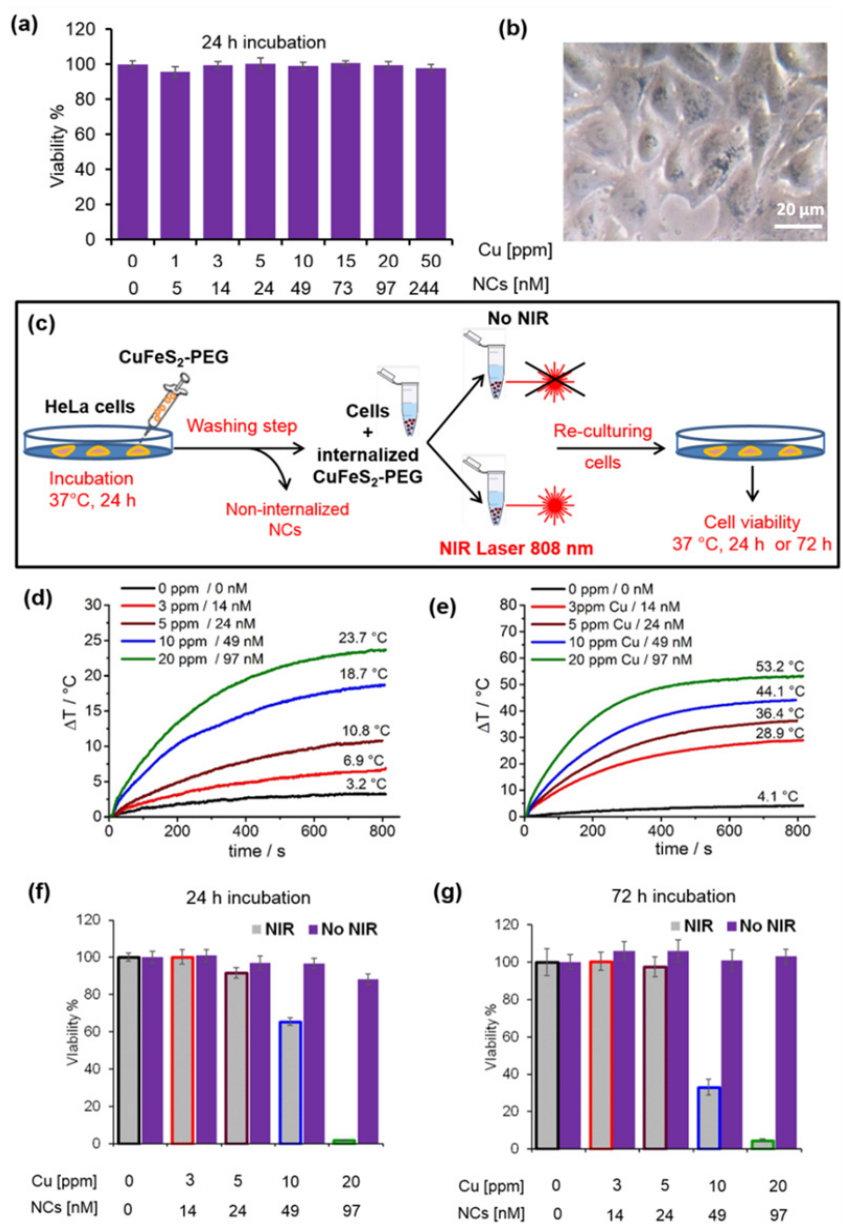

**Figure 3**. (a) Cell viability test after 24 h of incubation of HeLa cells with solutions of $CuFeS_2$–PEG NCs under radiationless conditions and at different NC concentrations. (b) Optical image of HeLa cells after exposure to NCs for 24 h at 37 °C, 5% $CO_2$ at [Cu] of 20 ppm - the bluish spots are internalized NCs. (c) Schematic summary of the cell *in-vitro* photo-thermal experiment. (d) Temperature profile under laser irradiation of HeLa cells in medium solution previously incubated at varying amounts of initial administered NCs for 24 h at 37 °C. (e) Temperature profile under laser irradiation of just NCs dissolved in cell medium at different concentrations and corresponding to the initial NC doses administered to the cells. The initial temperature for both (d) and (e) was around 30 °C. Experiment of laser irradiation effects on cells (f and g). HeLa cells were exposed for 24 hours to $CuFeS_2$-NCs; after washing the cells from the non-associated $CuFeS_2$–NCs, the cells collected in 0.5 ml medium were irradiated by the laser for 13 minutes (NIR) or they were kept at RT for the same time interval (No NIR). On the re-cultured cells, viability was assessed after (f) 24 h and (g) 72 h. By the comparison of cells exposed and not exposed to the NIR, it is clear that the irradiation-induced cytotoxicity increases in a dose dependent manner. The cell viability was measured using control cells (i.e. cells not incubated with NCs) as a reference.

$CuFeS_2$-PEG NCs, either in water or in complete DMEM, were stable at 37 °C even after 24 h, with no sign of aggregation or precipitation (the spectral characteristics and color of the NC solution did not change and no aggregation was observed at the bottom of the vials, see Figure S10). A similar test was also performed under laser illumination, to ascertain the colloidal stability of $CuFeS_2$-PEG NCs during heating cycles. A NC solution at 15 ppm [Cu] was placed inside a DLS cuvette and irradiated with the 808 nm laser for several cycles. DLS recorded after each irradiation cycle showed no significant changes, confirming the stability of these NCs under laser irradiation (see Figure S11 and section S8 for more details). The heating-cooling temperature profiles under laser irradiation were consistently reproduced over five consecutive cycles, showing the sample consistency under repetitive heating cycles (Figure S12).

**Cell studies.** An ideal photothermal agent should exhibit minimal intrinsic toxicity to the living cells under radiationless conditions. In order to clear this point in the present case, the biocompatibility of the $CuFeS_2$-PEG NCs was evaluated on

HeLa cells by means of a PB assay. Solutions at various concentrations of CuFeS$_2$-PEG NCs, in a range from 5 nM to 244 nM (corresponding to [Cu] ranging from 1 to 50 ppm), were administered to HeLa cells and the toxicity was evaluated after 24 and 72 h of incubation at 37 °C. The cell viability after 24 h was not affected by the presence of the NCs, as no sign of cytotoxicity was detected in the whole concentration range analyzed (viability values above 97% were recorded, Figure 3a), in line with results reported for other chalcogenide NCs.[51] The internalized NCs were discernible under the optical microscope at [Cu] feeds above 10 ppm (bluish spots in Figure 3b). However, at 72 h of continuous NC exposure, the viability decreased with increasing concentration of the NC: the viability was above 80% for [Cu] up to 5 ppm while significant toxicity was observed when the cells were continuously exposed to NC solutions with [Cu] higher than 10 ppm (Figure S13). A similar toxicity trend was also confirmed when testing the same NCs on another tumor cell line, as for instance the IGROV 1 cells (Figure S14). The low toxicity of the CuFeS$_2$-PEG NCs in the range from 3 to 10 ppm motivated us to study their photothermal activity as well (as schematically shown in Figure 3c). In an *in-vitro* test, cultured cells were first exposed to NC solutions at different [Cu] doses (3, 5, 10 and 20 ppm) for 24 h at 37 °C to enable NC internalization by the cells before performing the laser treatment. After rinsing off the non-internalized NCs, one fraction of the cells were detached and re-suspended in a fresh medium for exposure to NIR radiation (1.14 W, 3.1 W cm$^{-2}$, spot size 0.36 cm$^2$, 13 min irradiation time). The other fraction, used as control, was kept for the same time span as the irradiated fraction at room temperature (around 30 °C), but was not irradiated (this sample is indicated as "No NIR" in Figure 3c). Upon laser irradiation, a rise in the temperature of the whole cell sample solution was observed, even for initial [Cu] values as low as 5 ppm. For cells exposed to [Cu] doses above 5 ppm, the temperature increase was significantly higher than that of pure water irradiated with the same laser at equivalent power and time.

The temperature rise was much more pronounced at increasing amounts of NC exposure, suggesting a progressively higher NC internalization by the cells at higher exposure dose, as seen from Figure 3d. The maximum temperature reached on the cell samples was always lower than that of the solutions prepared by dissolving an equal amount of NCs in 0.5 ml of DMEM (Figure 3e). This indicates that the amount of NCs internalized by the cells is definitely lower than that of the initially administered dose and indeed only a small fraction is taken up by the cells. After laser irradiation, the cells were re-cultured and the viability was assessed after 24 and 72 h of the laser treatment. For initial [Cu] doses of 10 ppm or above, a surge in cytotoxicity for Hela cells could already be distinguished after 24 h on the irradiated samples when compared to the non-irradiated control sample, as shown in Figure 3f. After 72h, the rise in cytotoxicity was much more pronounced at all the NC doses tested, indicating a permanent cell damage rather than a bare acute toxic effect (Figure 3g). The control cells, containing internalized NCs but not irradiated, did not show signs of toxicity ("No NIR" in Figure 3g). Another set of control cells, irradiated with the laser but not incubated with the NCs, exhibited 100% viability. This stands as a proof that the laser light in the conditions of the experiments used (1.14 W, 3.1 W cm$^{-2}$ spot size 0.36 cm$^2$, 13 min irradiation time) did not affect the cell viability, unless it was partially absorbed by the NCs and was then converted into heat.

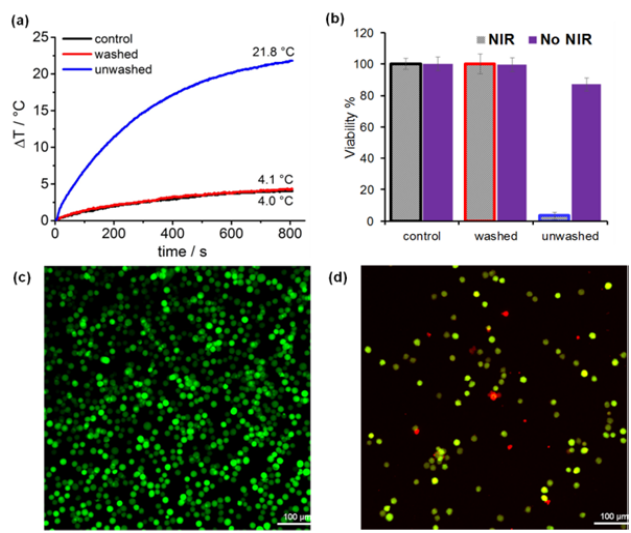

**Figure 4**. Identification of the minimal dose of CuFeS$_2$ NCs needed to produce cell damage only under laser irradiation. (a) Temperature profile under irradiation of cell samples incubated with CuFeS$_2$-PEG NCs (at a dose of 3 ppm [Cu] for 24 h), under irradiation. The red curve refers to the fraction of cells "washed" from non-internalized NCs prior to irradiation, whereas the blue curve refers to the cells that were "unwashed" from the NCs. The black line corresponds to the effect of irradiation of the control sample (cells not incubated with the NCs). In the unwashed sample, the non-internalized NCs contribute significantly to raise the cell temperature, which reached 50°C (the initial solution temperature was around 28 °C). (b) Cell viability recorded on the cells after laser treatment and after re-culturing for 72 h, indicating the high toxicity of the NCs in the case of "unwashed" sample. For each of the three cases, the purple histogram reports, as a control, the viability of cells treated in the same manner but not subjected to laser irradiation. Fluorescence confocal images of the irradiated cells (c) without, and (d) with NCs (at 3 ppm of [Cu]), after staining with calcein AM (green) and EthD-1 (red). The green color signifies healthy cells (as in c, without NCs), which decrease in number on irradiation in the presence of NCs with an accompanying rise in the number of damaged/dead cells signified by the presence of orange/red cells, as shown in (d).

In an *in vivo* photothermal treatment of a tumor, often the photothermal agents are intra-tumorally injected before exposing the tumor area to the laser treatment. This approach minimizes the amount of photosensitizer needed for an effective photo-thermal therapy, as it is delivered locally and not through the blood stream. A laser irradiation experiment, hence, was designed to mimic the case of an intra-tumoral injection of NCs by choosing the minimal dose of CuFeS$_2$–PEG NCs that is required to increase the temperature above 46 °C (which is considered high enough to induce tumor cell damage) and that at the same time does not manifest any sign of toxicity in dark conditions. In this connection, it is important to note that a minimal laser power of 3.1 Wcm$^{-2}$ was necessary to achieve equilibrium temperatures of around 46 °C (initial temperature being 30 °C, section S11). The temperature profiles (Figure 3e) and toxicity tests over prolonged exposure times (Figures S13 and S14), indicate that this dose corresponds to [Cu] of 3 ppm. A cell viability above 97% was recorded at this dose, even after 72 h of continuous NC exposure.

Upon irradiating a solution at this dose, the temperature rose from the initial 30 °C to 58.9 °C (30+28.9), as shown in Figure 3e.

After having identified 3 ppm in [Cu] as the optimal dose, we performed an *in vitro* experiment in which the cells were exposed to this dose of NCs for 24 h. In one case, the cells were not washed from the non-internalized NCs prior to irradiation with the laser. In another case, the cells underwent the same incubation conditions with the NCs, but they were rinsed before being irradiated. A control sample was represented by cells that were not incubated with NCs, but which were irradiated with the laser. Obviously, irradiation of the unwashed sample caused a much higher increase in temperature, from 28°C to 50 °C after 13 minutes exposure (Figure 4a, blue line), than in the washed sample (Figure 4a, red line) which behaves in practice as the control sample (Figure 4a, black line). The cells, after the different treatments, were re-cultured and their viability was assessed after 72 h. The viability of the "unwashed" sample was less than 5% (Figure 4 b, histogram marked with a blue line), while that of the "washed" sample was close to 100% (Figure 4 b, histogram marked with a red line), comparable to that of the control sample (Figure 4 b, histogram marked with a black line). For each of these three samples, we also ran a parallel experiment consisting in treating cells in the same manner but not subjecting them to laser irradiation. The viability for these parallel samples is reported in purple histograms in Figure 4b (No-NIR). In these latter samples, no toxicity effects were detectable. The overall conclusion from these experiments is that 3 ppm [Cu] (corresponding to 14 nM in NCs) represents a concentration that does not cause any toxicity effects on the cells, even after prolonged exposure, but it does become toxic under irradiation.

We performed an additional qualitative experiment to test cell viability after NIR irradiation, based on the simultaneous determination of live and dead cells through fluorescence imaging using calcein AM (green) and EthD-1 (red) as dyes, respectively, as presented in Figures 4c and 4d. Briefly, cells exposed to 3 ppm in [Cu] concentration of NC were first irradiated and then re-cultured to analyze their viability (after 5 h) under a fluorescence confocal microscope in comparison to cells irradiated without the exposure to the NCs. As expected, irradiated cells treated without the NCs were healthy and attached to the substrates (green color, Figure 4c) whereas most of the cells treated with 3 ppm in [Cu] dose, upon irradiation were dead (red color, Figure 4d) or were suffering (orange, Figure 4d). Furthermore, a lower number of irradiated cells exposed to the NCs were able to adhere to the multiwall (less fluorescent spots in Figure 4d than in Figure 4c).

CONCLUSIONS

We have described a colloidal synthesis of monodisperse and phase-pure chalcopyrite ($CuFeS_2$) NCs, which are characterized by an absorption spanning almost the entire visible and NIR regions, with broad maxima at 480 and 950 nm. The presence of an IB in the electronic band structure (as supported by DFT and TA spectroscopy) with a relatively small indirect VB-IB gap makes these NCs excellent candidates for light-to-heat conversion in the biological window of 650-900 nm. Consequently, chalcopyrite NCs exhibit enhanced PT efficiency of 49% under laser radiation of 808 nm. In addition, the small hydrodynamic diameters of these NCs coupled with high natural abundance of the chalcopyrite mineral (and hence of the metals Cu and Fe) makes them excellent candidates for photothermal therapy. Tumor cell annihilation upon laser irradiation in the presence of $CuFeS_2$ NCs demonstrate their therapeutic ability. The NCs are stable in biological liquids so that, in principle, they can be injected into organisms, either intra-tumorally or intravenously, which will constitute our future work.

It is evident from our studies that the otherwise detrimental effects of iron on photovoltaic applications can be a boon for applications like photothermal therapy. The same concept of introducing deep Fe intermediate levels in conventional semiconductors can be used to design a new class of photothermal agents, which operate through a different light to heat conversion mechanism than that for plasmonic NCs like gold and copper chalcogenides. Indeed, iron doping in other chalcopyrite type semiconductors (like $CuGaS_2$ and $CuAlS_2$) introduces similar effects as in $CuFeS_2$.

## ASSOCIATED CONTENT

**Supporting Information**. Additional experiments on NC growth monitoring, thermogravimetric analysis of $CuFeS_2$-PEG NCs, theoretical electronic band structure calculations, normalized TA spectra of the NCs, calculations of molar attenuation coefficient (ε) and PT efficiency, DLS data on NC stability in physiological conditions, colloidal stability of NCs under laser irradiation, reversibility of the heating-profile curves, and additional data on cell viability and photothermal experiments. This material is available free of charge *via* the Internet at http://pubs.acs.org.

## AUTHOR INFORMATION

**Corresponding Authors**

*E-mails: sandeep.ghosh@iit.it, teresa.pellegrino@iit.it, liberato.manna@iit.it

**Author Contributions**

‡ These authors contributed equally

## ACKNOWLEDGMENT

The research leading to these results has received funding from the FP7 under grant agreement n. 614897 (ERC Consolidator Grant "TRANS-NANO") and from the Horizon 2020 under grant agreement No 678109 (ERC starting grant "ICARO")

TOC Graphics

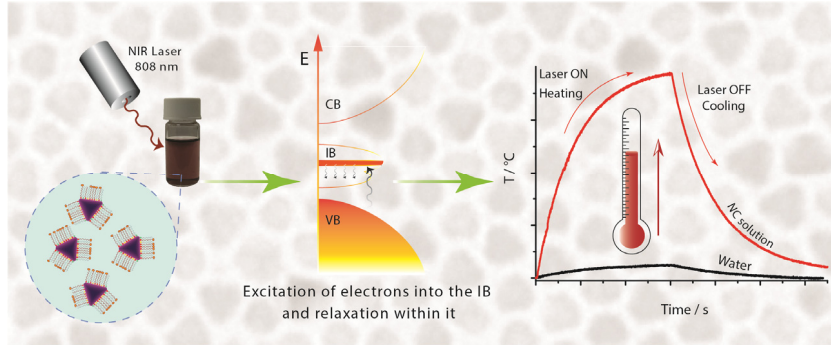